# Automatic Detection and Classification of Cognitive Distortions in Mental Health Text

Benjamin Shickel, Scott Siegel, Martin Heesacker, Sherry Benton, and Parisa Rashidi

**Abstract**—In cognitive psychology, automatic and self-reinforcing irrational thought patterns are known as cognitive distortions. Left unchecked, patients exhibiting these types of thoughts can become stuck in negative feedback loops of unhealthy thinking, leading to inaccurate perceptions of reality commonly associated with anxiety and depression. In this paper, we present a machine learning framework for the automatic detection and classification of 15 common cognitive distortions in two novel mental health free text datasets collected from both crowdsourcing and a real-world online therapy program. When differentiating between distorted and non-distorted passages, our model achieved a weighted F1 score of 0.88. For classifying distorted passages into one of 15 distortion categories, our model yielded weighted F1 scores of 0.68 in the larger crowdsourced dataset and 0.45 in the smaller online counseling dataset, both of which outperformed random baseline metrics by a large margin. For both tasks, we also identified the most discriminative words and phrases between classes to highlight common thematic elements for improving targeted and therapist-guided mental health treatment. Furthermore, we performed an exploratory analysis using unsupervised content-based clustering and topic modeling algorithms as first efforts towards a data-driven perspective on the thematic relationship between similar cognitive distortions traditionally deemed unique. Finally, we highlight the difficulties in applying mental health-based machine learning in a real-world setting and comment on the implications and benefits of our framework for improving automated delivery of therapeutic treatment in conjunction with traditional cognitive-behavioral therapy.

## I. INTRODUCTION

ACCORDING to the National Institute of Mental Health, anxiety disorders affect more than 18% of the U.S. adult population every year [1]. Additionally, the National Survey of Drug Use and Health reports that 6.7% of the U.S. adult population experienced at least one major depressive disorder episode in the past year [2]. Cognitive-behavioral therapy (CBT) has been shown to have a large positive effect on patients experiencing symptoms of anxiety and depression [2]. A large part of CBT treatment is helping the patient learn to self-identify his or her automatic and oftentimes irrational thought patterns that contribute to a distorted perception of reality. These negative thoughts are referred to as cognitive distortions, which have been grouped into 15 common types [3].

From a cognitive-behavioral perspective, cognitions can be viewed as the intermediary link between an external stimulus and a subjective feeling, taking the form of internalized statements a person tells themselves [4]. Cognitive distortions denote such self-statements that are either mildly misinterpreted or entirely inaccurate, often reflected by logical fallacies in internal thinking. For example, *"they didn't respond to my text message, they must be ignoring me"* is a cognitive distortion because it makes definitive assumptions about a third party that is impossible to know with certainty. In this example, a non-distorted cognition might be *"they didn't respond to my text message, maybe they are preoccupied or haven't seen it yet"*. In many cases, negative subjective feelings such as anxiety are responses to these types of distorted cognitions. Thus, a large focus of cognitive-behavioral therapy involves training patients to identify their own cognitive misjudgments to improve overall well-being. A full list of all 15 cognitive distortions used in our experiments, along with a short description of each, can be found in Table. A.I located in the Appendix.

In recent years, online therapy programs have been developed to supplement or replace traditional cognitive-behavioral therapy [5]–[12]. Similar to traditional CBT, these online counterparts provide guidance and instruction for identifying and challenging cognitive distortions and negative feedback loops. A key component of these online services is the inclusion of mental health journals and logs, where users describe their thoughts, emotions, anxiety events, and self-assessment.

In this paper, we develop methods for automatically detecting and classifying cognitive distortions in mental health text using machine learning techniques. Such frameworks have important implications for the delivery of effective online mental health services. Along with assisting traditional therapist intervention, automated distortion assessment can provide instant feedback and guidance to users exhibiting distorted thinking at any time of day. These frameworks would also allow for online therapy to scale to an even larger number of users. Additionally, these tools can provide early warning of more severe mental illness and can potentially be used to monitor social media for ideal candidates for mental health

This work was supported by NSF-IIP 1631871 from the National Science Foundation (NSF), Division of Industrial Innovation and Partnerships (IIP).

B. Shickel, S. Siegel, M. Heesacker, and P. Rashidi are with the University of Florida, Gainesville, FL 32611 USA (e-mail: shickelb@ufl.edu; sns08j@ufl.edu; heesack@ufl.edu, parisa.rashidi@ufl.edu).

S. Benton is with TAO Connect, Inc., St. Petersburg, FL 33701 USA (e-mail: sherry.benton@taoconnect.org).

services.

Our primary contributions can be summarized by the following:

- To our knowledge, we present the first attempts at detecting and classifying a large range of cognitive distortions (15) from text using machine learning techniques.
- We collect a novel dataset of real-world cognitive distortion event recollections from crowdsourcing participants.
- We develop a second novel dataset of online mental health therapy logs annotated by experts for presence of cognitive distortions.
- We examine a reduced set of clinically recognized cognitive distortions based on unsupervised machine learning techniques.

## II. Methods

This study was approved by the University of Florida Institutional Review Board IRB 201401089, and all methods were performed in accordance with the relevant guidelines and regulations. Informed consent was obtained from subjects before enrollment in the study.

### A. Data

Since cognitive distortion detection is a novel machine learning task, at present time there are no publicly available datasets containing text passages with labeled distortions. Such data are necessary for constructing supervised machine learning models suitable for predicting and classifying text-based distortions. To this end, we collected and annotated three novel cognitive distortion datasets, which we describe in this section. These datasets are summarized in Table. I. The CrowdDist and MH-C datasets are labeled with 15 cognitive distortions, while MH-D is annotated with binary distortion/no distortion labels.

*1) Crowdsourced distortion recollections (CrowdDist):* Our first cognitive distortions dataset, which we refer to as Crowd-Dist, comes from the popular crowdsourcing platform Mechanical Turk[1] (MTurk). For all assignments, we enforced a minimum quality requirement for MTurk workers of having at least 95% of their previous assignments accepted. Workers were compensated $1.00 for completing an assignment, which took 15.3 minutes on average to complete. To reduce annotation bias, individuals were only permitted to complete one assignment each.

When MTurk workers agreed to participate in the assignment, they were randomly assigned to one of the three groups, each of which contained a different set of five cognitive distortions. For each of the five distortions assigned to the worker, they were first presented with a short description of the types of thinking that the particular distortion exemplifies. Then, they were asked to tell us about an actual, specific time from their own life in which they exhibited the same type of described thinking. Workers were encouraged to provide approximately 2-3 sentences, but no limit was enforced. Distortion groups were randomly assigned to arrive at an even distribution of responses per distortion. In total, we collected 8,940 text passages from 1,788 unique individuals from MTurk's pool of qualified workers. After collection, these responses were manually reviewed for quality and relevance. After removing irrelevant contributions, the final dataset contained 7,666 text responses across all 15 distortions, with an average of 511 responses per distortion.

*2) Mental health therapy logs (MH):* Our second dataset, which we refer to as MH, comes from TAO Connect[2], an online mental health therapy service aimed at improving coping strategies for college students suffering from anxiety and depression. As part of treatment, patients completed regular journals describing their mental state, anxiety events, specific worries, and progress. We recruited and trained four senior undergraduate psychology students to provide cognitive distortion annotations for these mental health journals. Each annotator was instructed to select one or more cognitive distortions that each journal response exhibited, or to indicate that no distortions were exhibited.

From the overall MH dataset, we constructed two subsets for each of our two cognitive distortion tasks. The MH-C dataset was annotated with 15 cognitive distortion labels, and the MH-D dataset was annotated with binary distorted/non-distorted labels. The MH-C dataset was used in the cognitive distortion classification task, thus representing a multi-class classification problem. We only kept passages in which a majority of annotators agreed on which cognitive distortion was exemplified by the passage. For detecting presence of distorted text, we assigned a binary label to each passage indicating whether any cognitive distortion was selected by the annotators. We only kept passages in which a majority of annotators selected at least one cognitive distortion.

Below are a few short narratives taken from the MH-D dataset that exemplify the presence or absence of cognitive distortions in these mental health logs:

Examples of **non-distorted** passages:

- *"Thinking about life after school whether or not I'll be able to find a job."*
- *"I was feeling calm all day, no worries yesterday."*
- *"I was told the news that my grandfather had passed away, I was worried about how my family was going to cope."*

Examples of **distorted** passages:

- *"Exam tomorrow and missing someone thinking that I was going to fail and lose that person forever."*
- *"Forgot a meeting with my advisor, inconvenienced him – I'm incompetent."*
- *"My friend didn't sit next to me, I thought I was losing a friend."*

In the MH-D dataset, there exists a large class imbalance between distorted and non-distorted text (Table. I). The large class skew is a result of the source of our data being a mental health treatment service, and specifically the use of anxiety monitoring logs. Thus, we fully expect most responses to reflect emotional distress and contain some type of cognitive distortion.

TABLE I
SUMMARY STATISTICS FOR CROWDDIST AND MH DATASETS.

| | **CrowdDist** | **MH-C** | **MH-D** |
|---|---|---|---|
| **Task** | Classification | Classification | Detection |
| **Source** | Mechanical Turk | Mental health logs | Mental health logs |
| **Guided collection** | Yes | No | No |
| **Passages, n** | 7,666 | 1,164 | 1,799 |
| **Unique labels, n** | 15 | 15 | 2 |
| **Median passage length, words** | 47 | 43 | 42 |
| **Annotators per passage, n** | 1 | 4 | 4 |
| **Distortion prevalence, n (%)** | | | |
| Distorted | 7,666 (100.0) | 1,164 (100.0) | 1,605 (89.2) |
| Not Distorted | 0 (0.0) | 0 (0.0) | 194 (10.8) |
| **Distortion counts, n (%)** | | | |
| Being Right | 536 (7.0) | 0 (0.0) | —— |
| Blaming | 494 (6.4) | 23 (2.0) | —— |
| Catastrophizing | 545 (7.1) | 53 (4.6) | —— |
| Control Fallacy | 490 (6.4) | 60 (5.2) | —— |
| Emotional Reasoning | 500 (6.5) | 187 (16.1) | —— |
| Fallacy of Change | 499 (6.5) | 0 (0.0) | —— |
| Fallacy of Fairness | 495 (6.5) | 1 (0.1) | —— |
| Filtering | 545 (7.1) | 386 (33.2) | —— |
| Global Labeling | 493 (6.4) | 4 (0.3) | —— |
| Heaven's Reward Fallacy | 490 (6.4) | 0 (0.0) | —— |
| Mind Reading | 545 (7.1) | 260 (22.3) | —— |
| Overgeneralization | 546 (7.1) | 22 (1.9) | —— |
| Personalization | 497 (6.5) | 25 (2.1) | —— |
| Polarized Thinking | 497 (6.5) | 123 (10.6) | —— |
| Should's | 494 (6.4) | 20 (1.7) | —— |

*3) Preprocessing:* For both cognitive distortion tasks, we applied a sequence of common-practice techniques for processing natural language text before passing to our machine learning models for prediction. These steps included converting all passages to lowercase, removing punctuation, and tokenizing each processed passage based on white space delimiting.

We then used the generated word tokens to construct a collection of unigrams (single-words) and bigrams (two adjacent words) for every textual passage in our datasets. These n-grams were then transformed into sparse vectors in a bag-of-words fashion using these token counts, which we normalized by term frequency-inverse document frequency (tf-idf), a measure that provides information regarding each n-gram's relevance to individual documents compared with the overall corpus. When evaluating our models, we vectorized test data using counts of only the n-grams that occurred in the training data.

### B. Model

In this section, we report experimental settings and results for both the cognitive distortion detection and classification tasks.

We define cognitive distortion detection as the ability for a machine learning classifier to distinguish between text containing a distortion and text that is not distorted, and cognitive distortion classification as the ability for a model to distinguish between the 15 distortion categories in text that is already known to be distorted. Both tasks are important from a mental health perspective, as once text is determined to contain some distortion, knowing which distortion is present can guide treatment options for that patient. Because our CrowdDist dataset exclusively contains distorted text, we only report its distortion classification results.

For both classification tasks, we experimented with a variety of machine learning models including logistic regression, support vector machines (SVM), random forests, gradient boosted trees (XGBoost), recurrent neural networks (RNN), and convolutional neural networks (CNN). For simplicity, we only report results for the best-performing model: logistic regression. In contrast with more recent deep learning models in which interpretability is often sacrificed for predictive performance, logistic regression has the added benefit of being highly transparent, and in our datasets also resulted in superlative performance compared with more complex approaches. We hypothesize that logistic regression outperformed deep learning techniques for our particular tasks due to the relatively small size of our datasets, and the tendency of cognitive distortions to present themselves via specific, common emotional expressions that bag-of-words-based models more easily capture. We also attempted a transfer learning approach by training a model on our CrowdDist dataset and fine-tuning on the MH-C dataset, but this did not significantly improve results in either task.

We note in passing that aside from logistic regression, the second-best model for both cognitive distortion tasks was a bidirectional recurrent neural network with gated recurrent units (GRU) using pre-trained GloVe word embeddings. We refer interested readers to a more comprehensive overview of deep learning and word embedding techniques [13], [14].

We utilized a standard logistic regression model for both tasks: (1) predicting if a given text represents a distorted or non-distorted narrative, and (2) for narratives known to be distorted, classifying the text as exhibiting one of 15 potential cognitive distortions. Input features were extracted from the unigrams and bigrams of each narrative via term frequency-inverse document frequency (tf-idf) as outlined in the preprocessing section, and a separate logistic regression classifier was constructed for each of our two cognitive distortion classification tasks. Optimal hyperparameters were selected via grid search using a nested cross-validation procedure, and included model regularization, length of n-grams, and minimum and maximum document frequency thresholds for vectorizing words and phrases.

Similar to linear regression for continuous outputs, logistic regression uses a sigmoid activation and maximum likelihood estimation for categorical output labels to generate continuous scores ranging from 0 to 1, interpreted as the probability of a narrative belonging to each output class. For classifying narratives as containing cognitive distortions, a single logistic regression model was used. For classifying distorted text as one of 15 specific distortion labels, a separate logistic regression model was trained for all possible labels and aggregated in a "one vs. rest" approach. While a full overview of logistic regression is beyond the scope of this study, we refer interested readers to Hosmer, et al. for a more detailed overview of its technical aspects [15].

### C. Evaluation

When evaluating each individual dataset, we report the 5-fold nested cross-validation performance. When possible, we stratified the folds based on the distribution of distortions occurring in the dataset. In the MH-C dataset obtained from actual patients, some distortions resulted in very few annotated instances, so these instances were assigned to folds randomly. Since MH-C was annotated by multiple annotators, we only kept text responses in which the majority of annotators selected the same cognitive distortion, and discarded passages with no clear majority.

### D. Unsupervised exploration

In this study, we used the set of 15 cognitive distortions outlined in recent literature [16]; however, there is still no clear evidence-based consensus in the mental health and psychology communities on the best way to classify cognitive distortions. Various studies argue for lists ranging from five [17] to 50 [18] cognitive distortions. In this section, we detail our unsupervised exploratory analysis procedure using the text in both of our cognitive distortion datasets. For all unsupervised analysis, we use the tf-idf representations of each passage as feature vectors.

*1) Hierarchical clustering:* We examined both datasets from a hierarchical clustering perspective, with the hypothesis that there may exist natural and hierarchical groupings of cognitive

| Label | N | Precision | Recall | F1 |
|---|---|---|---|---|
| Not Distorted | 194 | 0.57 | 0.29 | 0.38 |
| Distorted | 1605 | 0.92 | 0.97 | 0.95 |
| **All Examples (Macro)** | **1799** | **0.75** | **0.63** | **0.66** |
| **All Examples (Weighted)** | **1799** | **0.88** | **0.90** | **0.88** |

distortions that share common traits. For improved visualization, we took the sum of all tf-idf representations for each of the 15 cognitive distortions, resulting in a single aggregate feature vector per distortion. Clustering was performed with Ward's method of decreasing cluster variance using cosine similarity as the measure of distance.

*2) Latent Dirichlet allocation:* We also used Latent Dirichlet Allocation (LDA), a popular unsupervised topic-modeling algorithm for extracting a set of thematic topics composed of distributions of words from a corpus of text documents. It is a probabilistic method which assumes each document is generated by sampling words from a distribution of topics, where each topic is a distribution over the words in the corpus. In our experiments, we found n = 25 topics to yield the most meaningful topics, and we use this setting for LDA-based analysis. Once the LDA model was trained on an entire corpus, each document was converted into a 25-dimensional probability distribution over topics. For each of the 15 cognitive distortions, we took the sum of all topic probabilities from each of the distortion's passages as the overall LDA representation of the given distortion. We compared the pairwise cosine similarity between each pair of distortions as a measure of how related two distortions were in terms of their thematic content, repeating this process for every pair of two cognitive distortions.

## III. RESULTS

In this section, we begin by reporting classification performance for both the cognitive distortion detection and classification tasks. We then describe the results of our unsupervised distortion analysis.

### A. Cognitive distortion detection

Our first task, in which we wish to predict whether a text passage contains evidence of a cognitive distortion or is non-distorted, only applies to our online therapy dataset. In the CrowdDist dataset, in which we directly elicited responses from participants matching each of the 15 cognitive distortions, we do not have text that is labeled as non-distorted.

The 5-fold cross validation classification experiments resulted in a weighted F1 score of 0.88 for all classes (Table. II), with the weighted precision, recall, and F1 score for the distorted class (0.92, 0.97, 0.95, respectively) greatly outperforming the same non-distorted classification metrics (0.57, 0.29, 0.38, respectively). Accuracy for this task was 90%, but given the significant class imbalance, we place more emphasis on F1 score for our primary classification metric.

Cognitive distortion classification results in CrowdDist and MH-C datasets.

| | **CrowdDist** | | | | **MH-C** | | | |
| Label | N | Precision | Recall | F1 | N | Precision | Recall | F1 |
|---|---|---|---|---|---|---|---|---|
| Being Right | 536 | 0.73 | 0.78 | 0.75 | 0 | —— | —— | —— |
| Blaming | 494 | 0.77 | 0.72 | 0.74 | 23 | 0.00 | 0.00 | 0.00 |
| Catastrophizing | 545 | 0.67 | 0.71 | 0.69 | 53 | 0.50 | 0.21 | 0.29 |
| Control Fallacy | 490 | 0.65 | 0.64 | 0.64 | 60 | 0.28 | 0.22 | 0.25 |
| Emotional Reasoning | 500 | 0.55 | 0.55 | 0.55 | 187 | 0.40 | 0.40 | 0.40 |
| Fallacy of Change | 499 | 0.74 | 0.75 | 0.75 | 0 | —— | —— | —— |
| Fallacy of Fairness | 495 | 0.81 | 0.73 | 0.77 | 1 | 0.00 | 0.00 | 0.00 |
| Filtering | 545 | 0.70 | 0.74 | 0.72 | 386 | 0.52 | 0.62 | 0.56 |
| Global Labeling | 493 | 0.61 | 0.53 | 0.57 | 4 | 0.00 | 0.00 | 0.00 |
| Heaven's Reward Fallacy | 490 | 0.65 | 0.67 | 0.66 | 0 | —— | —— | —— |
| Mind Reading | 545 | 0.61 | 0.67 | 0.64 | 260 | 0.50 | 0.60 | 0.55 |
| Overgeneralization | 546 | 0.64 | 0.63 | 0.63 | 22 | 0.43 | 0.14 | 0.21 |
| Personalization | 497 | 0.68 | 0.63 | 0.66 | 25 | 0.00 | 0.00 | 0.00 |
| Polarized Thinking | 497 | 0.61 | 0.60 | 0.61 | 123 | 0.39 | 0.37 | 0.38 |
| Should's | 494 | 0.73 | 0.77 | 0.75 | 20 | 0.50 | 0.05 | 0.09 |
| **All Examples (Macro)** | **7666** | **0.68** | **0.68** | **0.68** | **1164** | **0.29** | **0.22** | **0.23** |
| **All Examples (Weighted)** | **7666** | **0.68** | **0.68** | **0.68** | **1164** | **0.44** | **0.47** | **0.45** |

## B. Cognitive distortion classification

Our second task involved the classification of distorted text into the 15 distinct distortion categories shown in Table. I. For this task, each response is known to contain some unspecified cognitive distortion, whether by crowdsourced volunteers (CrowdDist) or annotated as such from online mental health journals (MH-C). Our aim is to predict which distortion the text contains. Specific, fine-grained identification of cognitive distortions is important for tailoring mental health treatment and providing specific feedback to both patients and therapists.

We note that the distribution of distortion labels is roughly balanced in the CrowdDist dataset (Table. I). This is a direct result of the guided nature of our data collection, where we intentionally collected a balanced distribution of responses for each of the 15 cognitive distortion categories. In contrast, the MH-C dataset comes from real-world patient journals in a currently active online mental health therapy service. The distribution of distortions in MH-C is much different, with some labels receiving zero annotator votes. While CrowdDist can be viewed as the ideal classification setting, MH-C exhibits the difficulties that come with modeling real-world data.

Table. III shows the precision, recall, and F1 scores for the cognitive distortion classification task in both datasets. The CrowdDist model yielded 0.68 accuracy and the MH-C model yielded 0.47 accuracy. Given 15 possible distortion labels, the accuracy represented by random chance is 0.06.

After obtaining the best models via 5-fold nested cross-validation, we retrained a single model per dataset using the optimal hyperparameters with all available data. Table. IV shows the ten most discriminative words and phrases for each of the 15 cognitive distortions in the CrowdDist dataset, found by taking maximum logistic regression term coefficients.

## C. Unsupervised exploration

Aside from constructing models to accurately predict and classify distorted text passages, we also sought to quanti-tatively suggest the natural number of unique distortions in a data-driven manner. For these experiments, we did not assume any fixed number of cognitive distortions, which is frequently debated in literature. Instead, we turned to unsupervised machine learning techniques to reveal natural cooccurrence and groupings of related text in both CrowdDist and MH-C passages containing cognitive distortions. For brevity, in this section we only show unsupervised analysis of the CrowdDist dataset, which is larger and more balanced than the MH dataset.

The result of hierarchical clustering on the distorted text passages from the CrowdDist dataset is shown as a dendrogram in Fig. 1. The clustering via Ward's method using cosine distance appears to suggest four natural groupings of cognitive distortions.

In Fig. 2, we examine the cosine similarity between every pair of our 15 cognitive distortions in the CrowdDist dataset, evaluated based on the sum of the LDA topic distribution for each passage of each cognitive distortion. In the CrowdDist dataset, the five most similar distortion pairs were Fallacy of Fairness/Heaven's Reward Fallacy (0.98), Emotional Reasoning/Mind Reading (0.98), Emotional Reasoning/Global Labeling (0.98), Being Right/Mind Reading (0.98), and Global Labeling/Mind Reading (0.95). In the MH-C dataset, the five most similar distortion pairs were Filtering/Polarized Thinking (0.98), Blaming/Filtering (0.97), Blaming/Emotional Reasoning (0.97), Catastrophizing/Filtering (0.96), and Blaming/Polarized Thinking (0.96).

## IV. Discussion

This study details the application of machine learning techniques toward detecting, classifying, and understanding the underlying structure of cognitive distortions in short text passages. There is a currently a substantial lack of annotated datasets in this domain, and one of our primary contributions is the collection of two novel cognitive distortion datasets

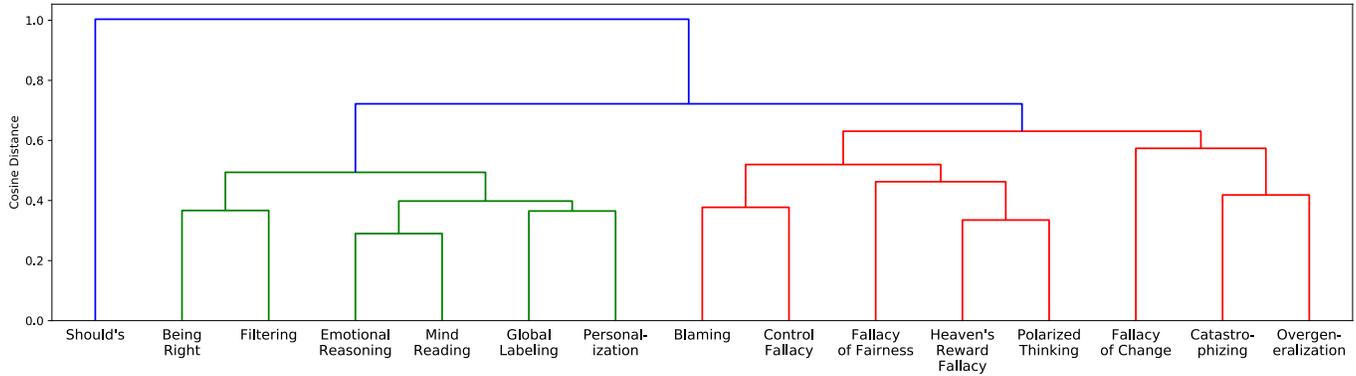

Fig. 1. Hierarchical clustering results for the CrowdDist dataset. Feature vectors for each of the 15 cognitive distortions were obtained by taking the sum of the tf-idf representations of passages for each label. Clusters obtained via Ward's method using cosine similarity as the measure of distance.

| | Being Right | Blaming | Catastrophizing | Control Fallacy | Emotional Reasoning | Fallacy of Change | Fallacy of Fairness | Filtering | Global Labeling | Heaven's Reward Fallacy | Mind Reading | Overgeneralization | Personalization | Polarized Thinking | Should's |
|---|---|---|---|---|---|---|---|---|---|---|---|---|---|---|---|
| Being Right | 1 | 0.46 | 0.58 | 0.43 | 0.59 | 0.36 | 0.55 | 0.54 | 0.74 | 0.38 | 0.71 | 0.61 | 0.58 | 0.54 | 0.57 |
| Blaming | 0.46 | 1 | 0.73 | 0.84 | 0.52 | 0.62 | 0.69 | 0.63 | 0.79 | 0.69 | 0.53 | 0.81 | 0.84 | 0.75 | 0.83 |
| Catastrophizing | 0.58 | 0.73 | 1 | 0.69 | 0.73 | 0.5 | 0.72 | 0.71 | 0.88 | 0.66 | 0.77 | 0.96 | 0.79 | 0.77 | 0.77 |
| Control Fallacy | 0.43 | 0.84 | 0.69 | 1 | 0.49 | 0.59 | 0.65 | 0.55 | 0.75 | 0.67 | 0.48 | 0.75 | 0.78 | 0.72 | 0.86 |
| Emotional Reasoning | 0.59 | 0.52 | 0.73 | 0.49 | 1 | 0.4 | 0.59 | 0.66 | 0.74 | 0.46 | 0.89 | 0.65 | 0.69 | 0.56 | 0.57 |
| Fallacy of Change | 0.36 | 0.62 | 0.5 | 0.59 | 0.4 | 1 | 0.39 | 0.4 | 0.52 | 0.43 | 0.41 | 0.53 | 0.61 | 0.45 | 0.55 |
| Fallacy of Fairness | 0.55 | 0.69 | 0.72 | 0.65 | 0.59 | 0.39 | 1 | 0.87 | 0.76 | 0.72 | 0.62 | 0.75 | 0.84 | 0.82 | 0.75 |
| Filtering | 0.54 | 0.63 | 0.71 | 0.55 | 0.66 | 0.4 | 0.87 | 1 | 0.72 | 0.6 | 0.72 | 0.69 | 0.81 | 0.72 | 0.64 |
| Global Labeling | 0.74 | 0.79 | 0.88 | 0.75 | 0.74 | 0.52 | 0.76 | 0.72 | 1 | 0.71 | 0.8 | 0.92 | 0.88 | 0.83 | 0.88 |
| Heaven's Reward Fallacy | 0.38 | 0.69 | 0.66 | 0.67 | 0.46 | 0.43 | 0.72 | 0.6 | 0.71 | 1 | 0.51 | 0.74 | 0.82 | 0.93 | 0.75 |
| Mind Reading | 0.71 | 0.53 | 0.77 | 0.48 | 0.89 | 0.41 | 0.62 | 0.72 | 0.8 | 0.51 | 1 | 0.71 | 0.74 | 0.63 | 0.61 |
| Overgeneralization | 0.61 | 0.81 | 0.96 | 0.75 | 0.65 | 0.53 | 0.75 | 0.69 | 0.92 | 0.74 | 0.71 | 1 | 0.84 | 0.83 | 0.84 |
| Personalization | 0.58 | 0.84 | 0.79 | 0.78 | 0.69 | 0.61 | 0.84 | 0.81 | 0.88 | 0.82 | 0.74 | 0.84 | 1 | 0.91 | 0.89 |
| Polarized Thinking | 0.54 | 0.75 | 0.77 | 0.72 | 0.56 | 0.45 | 0.82 | 0.72 | 0.83 | 0.93 | 0.63 | 0.83 | 0.91 | 1 | 0.83 |
| Should's | 0.57 | 0.83 | 0.77 | 0.86 | 0.57 | 0.55 | 0.75 | 0.64 | 0.88 | 0.75 | 0.61 | 0.84 | 0.89 | 0.83 | 1 |

Fig. 2. Cosine similarity between the sum of each labeled document's 25-topic LDA topic distribution in the CrowdDist dataset. Pairs of cognitive distortions that are more similar in thematic content are shown as darker squares.

| Being Right | Blaming | Catastrophizing | Control Fallacy | Emotional Reasoning |
|---|---|---|---|---|
| right | blame | what if | control | feel |
| was right | blamed | started | responsible | felt |
| correct | blaming | worry | responsible for | invited |
| wrong | blame my | die | help | me |
| prove | fault | panicked | of control | invite |
| explain | my | will | felt | guilty |
| point | blamed my | worried | to help | emotions |
| to prove | for not | what | feel responsible | go |
| feelings | for | panic | feel | but |
| arguing | because | going to | advice | feeling |

| Fallacy of Change | Fallacy of Fairness | Filtering | Global Labeling | Heaven's Reward Fallacy |
|---|---|---|---|---|
| change | unfair | negative | internal | cheated |
| happy | fair | great | external | worked |
| to change | unfair that | positive | met | hard |
| relationship | resentful | good | concluded | bitter |
| be happy | fair that | said | concluded that | shocked |
| if | was unfair | on | loser | deserved |
| happiness | pulled | but | women | put |
| would | got | comment | people | that if |
| him | people | focus | men | rewarded |
| happier | ticket | nice | decided | sacrifice |

| Mind Reading | Overgeneralization | Personalization | Polarized Thinking | Should's |
|---|---|---|---|---|
| assume | never | happy for | failure | should |
| assumed | again | compare | either | people should |
| convinced | always | jealous | perfect | that should |
| response | would never | why | if | should be |
| reply | will | my friend | failed | people |
| she | ever | friend | or | rules |
| respond | will always | friend got | score | should always |
| at me | now | compared | or nothing | believe |
| they | because | compare myself | perfectly | believed |
| read | i'll never | myself | goal | college |

coming from both crowdsourcing volunteers (CrowdDist) and real mental health patients (MH).

In our cognitive distortion detection task, we developed a logistic regression model to predict whether text passages from mental health self-monitoring logs contained evidence of at least one cognitive distortion. Our model performed generally well with a weighted F1 score of 0.88 across all passages (Table. II). However, while the classifier was fully capable of correctly identifying distorted text (0.92 precision, 0.97 recall, 0.95 F1), it struggled with non-distorted passages from the same domain (0.57 precision, 0.29 recall, 0.38 F1 score). One rationale for poor performance in non-distorted text is the class imbalance in the MH-D dataset: only 10.8% of all text passages contained zero cognitive distortions. Since MH-D comes from a real-world mental health service, it is a logical expectation that most user journals will involve some type of distorted thinking. Furthermore, the relatively small size of MH-D (1,799 passages) would limit the ability of fully developing a domain-dependent language model and differentiating between text of each distortion class with very high accuracy.

When classifying the primary cognitive distortion of text already known to be distorted, our model performed well in the CrowdDist dataset with a weighted F1 score across all passages of 0.68 (Table. III), representing a clear improvement over the random-chance baseline of 0.06 given the 15 possible prediction classes. F1 scores for individual distortions ranged from 0.55 (Emotional Reasoning) to 0.77 (Fallacy of Fairness). Each distortion was characterized by specific themes and word choices, demonstrated by the top word coefficients of the logistic regression model. We believe the identification of these discriminative words can have positive impact on the implementation of existing online therapeutic treatment by serving as either early warning signs for particular cognitive distortions, or by drawing increased attention from human therapists to direct a more personalized treatment plan. While classification performance was overall satisfactory in the CrowdDist dataset, we believe that the short nature of the passages and the potential presence of multiple distortions had a negative impact on classification. Since distortions are known to cooccur and several distortions are similar in thematic content, our unsupervised analysis was designed to provide justification for reducing the number of cognitive distortions.

Distortion classification performance in the MH-C dataset was fair, with an overall weighted F1 score across all passages of 0.45 (Table. III), ranging from 0 (Blaming, Fallacy of Fairness, Global Labeling, Personalization) to 0.56 (Filtering). Again, the baseline accuracy for classifying 15 class labels by a random-chance classifier is 0.06. The primary difficulty in MH-C was the lack of annotation quantity for five distortions (Being Right, Fallacy of Change, Fallacy of Fairness, Global Labeling, Heaven's Reward Fallacy). Given the small dataset size and these underrepresented distortions, the 15-class classification problem proved especially difficult in the mental health domain. While psychologists and other mental health experts have identified 15 distinct cognitive distortions, our results indicate that some distortions occur much more frequently than others in a real-world setting. Combined with our unsupervised analysis of distortion similarity, we feel it is important to revisit these distortion categories for possible adjustment based on real patient data.

Our unsupervised analysis provided three perspectives on distortion similarity that represent the first steps toward a data-driven rationale for revisiting the distinction between distortions, and for possible reduction in their overall number.

When performing hierarchical clustering on both datasets (CrowdDist shown in Fig. 1), each experiment produced different domain-dependent groupings of distortions. A qualitative analysis of these results would indicate four inherent clusters in the CrowdDist dataset and just two in the MH-C dataset. There was no trend in the groupings of specific distortions between datasets, suggesting that thematic content in each domain played a large role in the way each distortion was exemplified.

We compared the cosine similarity between topic distributions using latent Dirichlet allocation in an attempt to quantify the relatedness between each distortion pair's thematic content (Fig. 2). Each domain produced different thematic similarity between distortions, but here we note some patterns between domains; for example, in both datasets, Polarized Thinking is highly similar to Should's, and Emotional Reasoning is highly dissimilar to Fallacy of Change.

Above all, both supervised and unsupervised experiments involving cognitive distortions highlight the importance of source domain when dealing with free text passages. We have showed that distorted text takes on different forms when coming from third-party individuals (CrowdDist) or real patients currently experiencing distorted thinking (MH). These experiments indicate the nature of cognitive distortions themselves: what is clear to healthy individuals may be more difficult to understand for patients affected by certain types of distorted thinking.

### A. Limitations

One potential limitation of our crowdsourced CrowdDist dataset is the fact that we prompted volunteers with specific descriptions of distortions. It is possible that the types of responses we gathered based on these scripted prompts do not accurately represent the types of mental health text found "in the wild". This bias was exemplified by the comparatively lower classification performance in the MH-C dataset.

Additionally, we discarded several examples of distorted mental health text for the distortion classification task based on there being no clear majority in annotator votes. While simplifying the tasks, we are potentially missing out on valuable distortion information, especially when several cognitive distortions are co-occurring.

Finally, the size of our real-world mental health dataset (MH) is less than ideal. It is difficult to draw definitive conclusions when not all distortions were equally represented in the data, and future work in this area should primarily focus on robust data collection.

### B. Comparison with prior work

While machine learning and deep learning methods have been successfully applied to many natural language processing classification tasks in other domains, relatively few works have explored these techniques in the context of mental health. Within this space, researchers have focused on tasks such as emotional valence prediction in mental health journals [19], identifying characteristic language indicators of mental health in social media [20]–[22], sentiment analysis of suicide notes [23], [24], early warning indicators for poor mental health in online forums [25], and detecting and phenotyping lifelong disorders like dementia and autism [26], [27]. Less attention has been paid to applying machine learning techniques from a cognitive-behavioral perspective as we have done in this paper.

Cognitive distortion detection and classification share similarities with the task of text-based emotion recognition [17], [28]–[30]. To our knowledge, only [17] has explicitly dealt with the classification of cognitive distortions. However, [17] used a list of five distortions, compared with our expanded list of 15. Furthermore, they worked exclusively with crowd-sourced data. To our knowledge we are the first to classify cognitive distortions using real-world, unscripted mental health text from online therapy programs.

### C. Conclusions

Cognitive distortions put people at significant risk for engaging in and sustaining dysfunctional behaviors, and for producing needlessly dysphoric affective reactions to experiences. So, a machine learning tool for detecting cognitive distortions from research participant responses and, more importantly, patient responses represent a critically important advance in computer-assisted healthcare delivery. This tool can be used to alert clinicians about the presence and frequency of distorted thinking exhibited in clients' verbal behavior, whether that verbal behavior is text or voice-to-text. Clinicians can use this tool to assess whether distorted thinking is initially a feature of clients' presenting concerns and if so, focus treatment toward changing distorted cognitions. Over time, assessment of distorted cognitions can provide clinicians with important data regarding whether treatment is producing desired results and whether over time clients who may not have initially presented with cognitive distortions start to reveal distorted thinking in their verbal behavior as the therapeutic alliance is developed and strengthened. One of the most useful features of this tool is that it is noninvasive. That is, clients simply engage in the normal verbal behavior associated with treatment, whether that is computer assisted or traditional treatment, and the machine learning tool assesses for distortions without any additional assessment or intervention. A closely-related useful feature is that client verbal behavior is occurring as part of the natural course of treatment, which provides increased confidence that what clients are writing or saying may be a reasonably accurate rendition of the covert verbal behavior in which they engage when not in therapy.

Because all client verbal behavior in therapy can be reviewed using this tool, therapists can be alerted to changes in distorted thinking exhibited from one session to the next or across multiple sessions. If distorted thinking steadily increases over time, alerting the therapist can allow for a refocusing of treatment or the use of different therapeutic interventions to address cognitive distortions. If distorted thinking steadily decreases over time, therapists can reasonably assume that treatment is working effectively in this domain and may wish to celebrate that progress with the patient. If the frequency of distorted thinking drops to the level of rarely occurring, therapists can shift the focus from treatment to maintenance of treatment gains and if other aspects of the client's functioning warrant it, the conclusion of formal treatment.

Accurately typing specific cognitive distortions and determining how many discrete distortions are present in a particular piece of client verbal behavior remain challenges. What we noticed in patient data that was absent in research participant data (as a direct result of the instructions to research participants) were cascades of cognitive distortions, which are cases where two, three, four or more cases of distorted thinking were exhibited in a single verbal response. Disentangling the pieces of these cascades is challenging. We hope to improve on the tool's ability to do that in the future. A closely related challenge involves improving on the accuracy of distortion typology. Some scholars have identified 10 distinct distortions, others 15, and some others have provided other estimates. Comparison of these lists reveals both the expected overlap between lists, but also often overlap within lists. Some elements on some lists appear to represent superordinate categories while what appear to be subordinate elements appear as other elements in the same list. All of this suggests that in the future, factor analytic and other statistical techniques for grouping distortions need to be employed to provide empirical clarity regarding the number of independent (or interdependent) dimensions of distortion that most accurately characterizes clients' cognitively distorted verbal behavior. This work will also result in more clarity about features that define any given cognitive distortion.

## Acknowledgment

Research supported by NSF-IIP 1631871 from the National Science Foundation (NSF), Division of Industrial Innovation and Partnerships (IIP). We thank TAO Connect for access and assistance with retrieving online therapy logs. The Titan X Pascal used for this work was donated by the NVIDIA Corporation through the GPU Grant Program.

# Appendix

| Distortion | Description |
|---|---|
| Being Right | We are continually on trial to prove that our opinions and actions are correct. Being wrong is unthinkable and we will go to any length to demonstrate our rightness. For example, "I don't care how badly arguing with me makes you feel, I'm going to win this argument no matter what because I'm right." Being right often is more important than the feelings of others around a person who engages in this cognitive distortion, even loved ones. |
| Blaming | We hold other people responsible for our pain, or take the other track and blame ourselves for every problem. For example, "Stop making me feel bad about myself!" Nobody can "make" us feel any particular way — only we have control over our own emotions and emotional reactions. |
| Catastrophizing | We expect disaster to strike, no matter what. This is also referred to as "magnifying or minimizing." We hear about a problem and use what if questions (e.g., "What if tragedy strikes?" "What if it happens to me?"). For example, a person might exaggerate the importance of insignificant events (such as their mistake, or someone else's achievement). Or they may inappropriately shrink the magnitude of significant events until they appear tiny (for example, a person's own desirable qualities or someone else's imperfections). |
| Control Fallacy | If we feel externally controlled, we see ourselves as helpless a victim of fate. For example, "I can't help it if the quality of the work is poor, my boss demanded I work overtime on it." The fallacy of internal control has us assuming responsibility for the pain and happiness of everyone around us. For example, "Why aren't you happy? Is it because of something I did?" |
| Emotional Reasoning | We believe that what we feel must be true automatically. If we feel stupid and boring, then we must be stupid and boring. You assume that your unhealthy emotions reflect he way things really are — "I feel it, therefore it must be true." |
| Fallacy of Change | We expect that other people will change to suit us if we just pressure or cajole them enough. We need to change people because our hopes for happiness seem to depend entirely on them. |
| Fallacy of Fairness | We feel resentful because we think we know what is fair, but other people won't agree with us. As our parents tell us when we're growing up and something doesn't go our way, "Life isn't always fair." People who go through life applying a measuring ruler against every situation judging its "fairness" will often feel badly and negative because of it. Because life isn't "fair" — things will not always work out in your favor, even when you think they should. |
| Filtering | We take the negative details and magnify them while filtering out all positive aspects of a situation. For instance, a person may pick out a single, unpleasant detail and dwell on it exclusively so that their vision of reality becomes darkened or distorted. |

**Continued on next page**

| Distortion | Description |
|---|---|
| Global Labeling | We generalize one or two qualities into a negative global judgment. These are extreme forms of generalizing, and are also referred to as "labeling" and "mislabeling." Instead of describing an error in context of a specific situation, a person will attach an unhealthy label to themselves. For example, they may say, "I'm a loser" in a situation where they failed at a specific task. When someone else's behavior rubs a person the wrong way, they may attach an unhealthy label to him, such as "He's a real jerk." Mislabeling involves describing an event with language that is highly colored and emotionally loaded. For example, instead of saying someone drops her children off at daycare every day, a person who is mislabeling might say that "she abandons her children to strangers." |
| Heaven's Reward Fallacy | We expect our sacrifice and self-denial to pay off, as if someone is keeping score. We feel bitter when the reward doesn't come. |
| Mind Reading | Without individuals saying so, we know what they are feeling and why they act the way they do. In particular, we are able to determine how people are feeling toward us. For example, a person may conclude that someone is reacting negatively toward them but doesn't actually bother to find out if they are correct. Another example is a person may anticipate that things will turn out badly, and will feel convinced that their prediction is already an established fact. |
| Overgeneralization | In this cognitive distortion, we come to a general conclusion based on a single incident or a single piece of evidence. If something bad happens only once, we expect it to happen over and over again. A person may see a single, unpleasant event as part of a never-ending pattern of defeat. |
| Personalization | Personalization is a distortion where a person believes that everything others do or say is some kind of direct, personal reaction to the person. We also compare ourselves to others trying to determine who is smarter, better looking, etc. A person engaging in personalization may also see themselves as the cause of some unhealthy external event that they were not responsible for. For example, "We were late to the dinner party and caused the hostess to overcook the meal. If I had only pushed my husband to leave on time, this wouldn't have happened." |
| Polarized Thinking | In polarized thinking, things are either "black-or-white." We have to be perfect or we're a failure — there is no middle ground. You place people or situations in "either/or" categories, with no shades of gray or allowing for the complexity of most people and situations. If your performance falls short of perfect, you see yourself as a total failure. |
| Should's | We have a list of ironclad rules about how others and we should behave. People who break the rules make us angry, and we feel guilty when we violate these rules. A person may often believe they are trying to motivate themselves with shoulds and shouldn'ts, as if they have to be punished before they can do anything. For example, "I really should exercise. I shouldn't be so lazy." Musts and oughts are also offenders. The emotional consequence is guilt. When a person directs should statements toward others, they often feel anger, frustration and resentment. |